\documentclass[11pt,titlepage]{article}
\renewcommand{\arraystretch}{1.5}
\input amssym.def
\input amssym.tex

\begin{document}

\pagestyle{plain}
\pagenumbering{arabic}
\def\thesection{\Roman{section}.}
\def\thesubsection{\Roman{section}-\Alph{subsection}.}
\def\theequation{\arabic{equation}}
\def\<{\begin{equation}}
\def\>{\end{equation}}
\def\build#1#2{\mathrel{\mathop{\kern 0pt#1}\limits_{#2}}}
\def\rf#1{(\ref{#1})}
\def\rhoh{\widehat{\rho}}
\def\qora{\overrightarrow{q}}
\def\pora{\overrightarrow{p}}
\def\qp{\qora,\pora}
\def\sgn{\mathop{\rm sgn}\nolimits}
\def\intd{\displaystyle\int}
\def\P{{\cal P}}
\def\Ph{\widehat{\P}}
\def\R{\mbox{\Bbb R}}
\def\kh{\widehat{\chi}}
\def\e#1{\mbox{e}^{#1}}
\font\Bbb=msbm10
\font\Bxsl=cmbxsl11

\renewcommand{\thefootnote}{\fnsymbol{footnote}}
\title{\hfill TIFR/TH/02-18 \\
\hfill PM/02-14 \\
\vfill
\LARGE\bf Bell inequalities in phase space and their violation
in quantum mechanics\thanks{Work supported by the
Indo-French Centre for the Promotion of Advanced Research, Project Nb
1501-02.}.}
\author{{\Large G. Auberson}\thanks{e-mail: auberson@lpm.univ-monpt2.fr} \\
	\sl Laboratoire de Physique Math\'ematique, UMR 5825-CNRS, \\
	\sl Universit\'e Montpellier II, \\
	\sl F-34095 Montpellier, Cedex 05, FRANCE.
\and
	{\Large G. Mahoux}\thanks{e-mail: mahoux@spht.saclay.cea.fr} \\
	\sl Service de Physique Th\'eorique, \\
	\sl Centre d'\'Etudes Nucl\'eaires de Saclay, \\
	\sl F-91191 Gif-sur-Yvette Cedex, FRANCE.
\and
	{\Large S.M. Roy}\thanks{e-mail: shasanka@theory.tifr.res.in} and
	{\Large Virendra Singh}\thanks{e-mail: vsingh@theory.tifr.res.in} \\
	\sl Department of Theoretical Physics, \\
	\sl Tata Institute of Fundamental Research, \\
	\sl Homi Bhabha Road, Mumbai 400 005, INDIA.
}
\date{May 23, 2002.}
\maketitle
\renewcommand{\thefootnote}{\arabic{footnote}}

\begin{abstract}
We derive ``Bell inequalities'' in four dimensional phase space and
prove the following ``three marginal theorem'' for phase space
densities $\rho(\qp)$, thus settling a long standing conjecture :
``there exist quantum states for which more than three of the quantum
probability distributions for $(q_1,q_2)$, $(p_1,p_2)$, $(q_1,p_2)$
and $(p_1,q_2)$ cannot be reproduced as marginals of a positive
$\rho(\qp)$''. We also construct the most general positive $\rho(\qp)$
which reproduces any three of the above quantum probability densities
for arbitrary quantum states. This is crucial for the construction of
a maximally realistic quantum theory.
\end{abstract}

\section{Joint probabilities of conjugate observables}
\setcounter{equation}{0}

A quantum system permits many different choices $\{A\}$, $\{B\}$,
$\{C\}$ \ldots for a complete commuting set (CCS) of observables. If
$\{\alpha\}$ denotes a set of eigenvalues of $\{A\}$, $\{\beta\}$ of
$\{B\}$,  $\{\gamma\}$ of $\{C\}$ etc, quantum theory predicts the
probabilities of observing $\{\alpha\}$ in the experimental context to
measure $\{A\}$, and similarly of $\{\beta\}$, $\{\gamma\}$ \ldots in
different contexts, but not a joint probability of $\{\alpha\}$,
$\{\beta\}$, $\{\gamma\}$ \ldots, because they refer to noncommuting
observables. Thus quantum theory predicts probabilities for observing
``eigenvalues'' $\{\overrightarrow{q}\}$ of position operators
$\overrightarrow{Q}$ in one context, or $\{\overrightarrow{p}\}$ of
momentum operators $\overrightarrow{P}$ in another context, but not
their joint probability.

We can ask whether the ``contextual'' quantum probabilities can be
extended so as to encompass joint probabilities of non commuting
observables. The difficulty in extending quantum probabilities for
different CCS of observables $\{A\}$, $\{B\}$, $\{C\}$ \ldots (which
do not mutually commute) to a joint probability of the different CCS
is the lesson learnt from decades of work on quantum contextuality
theorems \cite{GKS2}-\cite{Martin-Roy}. Of these the most celebrated
is Bell's theorem \cite{BCHSH} where the Einstein-locality postulate
in the context of the EPR paradox is equivalent to a postulate of
existence of a joint probability for the different CCS
$\{\overrightarrow{\sigma}_1.\overrightarrow{a},
\overrightarrow{\sigma}_2.\overrightarrow{b}\}$, 
$\{\overrightarrow{\sigma}_1.\overrightarrow{a},
\overrightarrow{\sigma}_2.\overrightarrow{b}'\}$, 
$\{\overrightarrow{\sigma}_1.\overrightarrow{a}',
\overrightarrow{\sigma}_2.\overrightarrow{b}\}$ and 
$\{\overrightarrow{\sigma}_1.\overrightarrow{a}'
\overrightarrow{\sigma}_2.\overrightarrow{b}'\}$
for the system of two spin-half particles. Here,
$\overrightarrow{\sigma}_1$ and $\overrightarrow{\sigma}_2$ are Pauli
spin operators for the two particles, $\overrightarrow{a}$,
$\overrightarrow{a}'$, $\overrightarrow{b}$, $\overrightarrow{b}'$ are
arbitrary unit vectors. The postulate leads to the Bell-CHSH \cite{BCHSH}
inequalities which are in conflict with quantum spin correlations.

Consider now the conjugate observables position and momentum. The
first phase space formulation of quantum mechanics is due to Wigner
\cite{Wigner} who defined the phase space distribution $\rho(\qp)$ to be
\[ \rho_W(\qp) \equiv \int{d\qora'\over{(2\pi)^{3N}}}\,
   \langle\qora-{\qora'\over2}|\,\rhoh\,|\qora+{\qora'\over2}\rangle\,
   \exp(\mbox{i}\pora.\qora')
\]
where $\rhoh$ is the density operator of the quantum state. The
marginals of this phase space distribution reproduce the quantum
position and momentum probabilities
\< \int d\pora\,\rho(\qp) = \langle\qora|\,\rhoh\,|\qora\rangle\,,\qquad
   \int d\qora\,\rho(\qp) = \langle\pora|\,\rhoh\,|\pora\rangle\,.
   \label{1}
\>
However, $\rho_W(\qp)$ cannot be interpreted as a phase space
probability density as it is not in general positive. Cohen and
Zaparovanny \cite{CZ} found the most general positive phase space
density function obeying the quantum marginal conditions \rf{1} for
configuration space dimension $N=1$, and Cohen \cite{Cohen} found them for
general $N$. These density functions can be considered as
generalizations of the simple uncorrelated positive function
$\langle\qora|\,\rhoh\,|\qora\rangle\langle\pora|\,\rhoh\,|\pora\rangle$
which satisfies \rf{1}. These results might raise hopes that quantum
probabilities of all CCS can be reproduced as marginals of one phase
space density. This is false.

Martin and Roy \cite{Martin-Roy} showed that for 2-dimensional configuration
space the postulate of existence of a positive phase space density is
in conflict with the hypothesis that appropriate ``marginals'' of this
density reproduce the quantum probability densities of the different
CCS $\{Q_1\cos\alpha+Q_2\sin\alpha,-P_1\sin\alpha+P_2\cos\alpha\}$ for
{\sl all} $\alpha$.

On the positive side, Roy and Singh \cite{RS1} discovered that not only the
quantum probabilities of the two CCS $\overrightarrow{Q}$ and
$\overrightarrow{P}$ (which have no observables in common) but in fact
the quantum probabilities for a chain of $(N+1)$ CCS, for example
$(Q_1,Q_2,\ldots,Q_N)$, $(P_1,Q_2,\ldots,Q_N)$, 
$(P_1,P_2,Q_3,\ldots,Q_N)$, \ldots $(P_1,P_2,\ldots,P_N)$ can be
simultaneously reproduced as marginals of one positive phase space density:
\<\begin{array}{rcl}
   \int dp_1dp_2\ldots dp_N\,\rho(\qp) &=&
      \langle q_1q_2\ldots q_N|\,\rhoh\,|q_1q_2\ldots q_N\rangle\,, \\
   \int dq_1dp_2\ldots dp_N\,\rho(\qp) &=&
      \langle p_1q_2\ldots q_N|\,\rhoh\,|p_1q_2\ldots q_N\rangle\,, \\
   &\vdots& \\
   \int dq_1dq_2\ldots dq_N\,\rho(\qp) &=&
      \langle p_1p_2\ldots p_N|\,\rhoh\,|p_1p_2\ldots p_N\rangle\,,
  \end{array} \label{2}
\>
They conjectured that it is impossible to find, for every quantum
state, a positive phase space density whose marginals reproduce
quantum probabilities of more than $N+1$ CCS of observables.

Our main purpose here is : i) to prove this long standing conjecture
and thus quantify the extent of simultaneous realizability of
noncommuting CCS as marginals of a positive phase space density, ii)
to construct explicitly the most general positive phase space density
which reproduces the quantum probabilities of the maximum number of CCS as
marginals. This will enable the construction of the most general
``maximally realistic'' quantum theory, generalizing the special
construction of Roy and Singh \cite{RS1}-\cite{RS2} which reproduces
$(N+1)$ CCS. The earliest realistic quantum theory, viz. that of
de~Broglie and Bohm \cite{dBB} (dBB) which reproduces only one CCS
(position) is of course not maximally realistic. It would be
interesting to compare particle trajectories of maximally realistic
quantum theories given here with the dBB trajectories.

From the mathematical standpoint, our basic results are theorems
concerning multidimensional Fourier transforms. They can be expected
to open up new applications in classical signal and image processing
as they vastly improve the earlier results of Cohen \cite{Cohen}
(which only considered marginals with no variables in common).

In the present paper, we restrict ourselves to the case $N=2$. We
formulate the four marginal problem and develop the new tool of
``phase space Bell inequalities'' to solve the problem. The resulting
``three marginal theorem'' proves the conjecture of Roy and Singh for
$N=2$. The most general maximally realistic phase space densities
reproducing quantum probabilities of three CCS are then explicitely
constructed. Full details are given in \cite{AMRS4}. The corresponding
results for general N, more involved, will be reported separately \cite{AMRS2N}.

\section{Four marginal problem}

Consider a four dimensional phase space with position variables
$(q_1,q_2)$ and momentum variables $(p_1,p_2)$. Let
$\{\sigma_{qq}(q_1,q_2),$ $\sigma_{qp}(q_1,p_2),$
$\sigma_{pq}(p_1,q_2),$ $\sigma_{pp}(p_1,p_2)\}$ be arbitrary given
normalized probability distributions. Is it possible to find a
normalized phase space density $\rho(\qp)$ of which $\sigma_{qq},$
$\sigma_{qp},$ $\sigma_{pq}$ and $\sigma_{pp}$ are marginals? i.e.
\<\begin{array}{c}
  \begin{array}{rclrcl}
   \int dp_1dp_2\,\rho(\qp) &=& \sigma_{qq}(q_1,q_2)\,, & \quad
   \int dp_1dq_2\,\rho(\qp) &=& \sigma_{qp}(q_1,p_2)\,, \\
   \int dq_1dp_2\,\rho(\qp) &=& \sigma_{pq}(p_1,q_2)\,, & \quad
   \int dq_1dq_2\,\rho(\qp) &=& \sigma_{pp}(p_1,p_2)\,,
  \end{array} \\
   \rho(\qp) \geq 0\,, \qquad
   \int d\qora d\pora\,\rho(\qp) = 1\,.
\end{array} \label{3}
\>
It is obvious from \rf{3} that the given probabilities $\sigma_{qq},$
$\sigma_{qp},$ $\sigma_{pq}$ and $\sigma_{pp}$ must at least obey the
consistency conditions
\< \sigma_{qq}, \sigma_{qp}, \sigma_{pq}, \sigma_{pp} \geq 0\,,
   \label{4}
\>
and
\<\begin{array}{rclrcl}
   \int dq_2\,\sigma_{qq}(q_1,q_2) &=& \int dp_2\,\sigma_{qp}(q_1,p_2)\,,
   & \quad
   \int dq_1\,\sigma_{qq}(q_1,q_2) &=& \int dp_1\,\sigma_{pq}(p_1,q_2)\,,\\
   \int dq_1\,\sigma_{qp}(q_1,p_2) &=& \int dp_1\,\sigma_{pp}(p_1,p_2)\,,
   & \quad
   \int dq_2\,\sigma_{pq}(p_1,q_2) &=& \int dp_2\,\sigma_{pp}(p_1,p_2)\,,
\end{array} \label{5}
\>
We therefore pose the following problem which we shall call {\bf the
four marginal problem} : {\sl Given four normalized probability
distributions $\sigma_{qq},$ $\sigma_{qp},$ $\sigma_{pq}$ and $\sigma_{pp}$ 
obeying the consistency conditions \rf{4} and \rf{5}, does there exist
any positive normalized phase space probability density $\rho(\qp)$
with these distributions as marginals?} Further, the special case
where the four given $\sigma$'s are quantum probability distributions
for eigenvalues of the corresponding CCS of observables will be of
great interest, and we shall call it {\bf the quantum four marginal
problem}. This means that the given probability distributions are of the form
\<\begin{array}{rclrcl}
  \sigma_{qq}(q_1,q_2) &=& |\langle q_1,q_2|\psi\rangle|^2\,, &\quad
  \sigma_{qp}(q_1,p_2) &=& |\langle q_1,p_2|\psi\rangle|^2\,, \\
  \sigma_{pq}(p_1,q_2) &=& |\langle p_1,q_2|\psi\rangle|^2\,, &\quad
  \sigma_{pp}(p_1,p_2) &=& |\langle p_1,p_2|\psi\rangle|^2\,,
\end{array} \label{6}
\>
for a pure quantum state $|\psi\rangle$, or of the analogous form
obtained by replacing $|\langle\xi|\psi\rangle|^2$ by
$\langle\xi|\,\rhoh\,|\xi\rangle$ for a quantum state with density
operator $\rhoh$. In this case, the consistency conditions are
automatically satisfied. A positive answer to the quantum four
marginal problem would imply simultaneous realizability of the four
CCS $(Q_1,Q_2)$, $(Q_1,P_2)$, $(P_1,Q_2)$ and $(P_1,P_2)$. We shall
see that in fact at most three CCS can be simultaneously realized.

\section{Phase space Bell inequalities}
Consider the functions $r(q_1,q_2)$, $s(q_1,p_2)$, $t(p_1,q_2)$ and
$u(p_1,p_2)$, defined by
\<\begin{array}{rclrcl}
  r(q_1,q_2) & = & \sgn F_1(q_1)\,\sgn F_2(q_2)\,, &\quad
  s(q_1,p_2) & = & \sgn F_1(q_1)\,\sgn G_2(p_2)\,, \\
  t(p_1,q_2) & = & \sgn G_1(p_1)\,\sgn F_2(q_2)\,, &\quad
  u(p_1,p_2) & = & -\sgn G_1(p_1)\,\sgn G_2(p_2)\,,
\end{array} \label{7}
\>
where $F_1$, $F_2$, $G_1$ and $G_2$ are arbitrary nonvanishing
functions. Then, it is obvious that
\<  r(q_1,q_2)+s(q_1,p_2)+t(p_1,q_2)+u(p_1,p_2) = \pm2 \qquad
   (\forall q_1,q_2,p_1,p_2). \label{8}
\>
Given four probability distributions obeying the consistency
conditions \rf{4} and \rf{5}, suppose that a normalized phase space
density $\rho(\qp)$ satisfying the four marginal conditions \rf{3}
exists. Multiplying eq.\rf{8} by $\rho(\qp)$ and integrating over
phase space, we deduce the phase space Bell inequalities
\< |S|\leq2\,, \label{9}
\>
where
\<\begin{array}{rcl}
  S &\equiv&  \intd dq_1dq_2\,r(q_1,q_2)\,\sigma_{qq}(q_1,q_2)
      +\intd dq_1dp_2\,s(q_1,p_2)\,\sigma_{qp}(q_1,p_2) \\
  & & +\intd dp_1dq_2\,t(p_1,q_2)\,\sigma_{pq}(p_1,q_2)
      +\intd dp_1dp_2\,u(p_1,p_2)\,\sigma_{pp}(p_1,p_2)\,.
\end{array} \label{10} \>
The necessary conditions \rf{9}-\rf{10} provide us with a proof that
the four marginal problem does not always admit a solution : choose the
probability distributions
\<\begin{array}{rcl}
  \sigma_{qq}(q_1,q_2) & = & {1\over2}\left[\delta(q_1-a_1)\delta(q_2-a_2)+
    \delta(q_1-a'_1)\delta(q_2-a'_2)\right]\,, \\
  \sigma_{qp}(q_1,p_2) & = & {1\over2}\left[\delta(q_1-a_1)\delta(p_2-b_2)+
    \delta(q_1-a'_1)\delta(p_2-b'_2)\right]\,, \\
  \sigma_{pq}(p_1,q_2) & = & {1\over2}\left[\delta(p_1-b_1)\delta(q_2-a_2)+
    \delta(p_1-b'_1)\delta(q_2-a'_2)\right]\,, \\
  \sigma_{pp}(p_1,p_2) & = & {1\over2}\left[\delta(p_1-b_1)\delta(p_2-b'_2)+
    \delta(p_1-b'_1)\delta(p_2-b_2)\right]\,.
\end{array} \label{11}
\>
Inequality \rf{9} is violated for functions $F$'s and $G$'s such that
\[ F_1(a_1),F_2(a_2),G_1(b_1),G_2(b_2) > 0\,, \quad
  F_1(a'_1),F_2(a'_2),G_1(b'_1),G_2(b'_2) < 0\,,
\]
which yields $S=4$.

To show that the Bell inequalities can be violated also in the
quantum case, and to find by how much, is not a trivial matter.

\section{Violation of phase space Bell inequalities in quantum theory}
Suppose next that the given probability distributions $\sigma_{qq},$
$\sigma_{qp},$ $\sigma_{pq}$ and $\sigma_{pp}$ are of the form \rf{6}
or of the corresponding forms in terms of an operator $\rhoh$. Notice
first that $\chi_1(q_1) \equiv {1\over2} \left [1+\sgn F_1(q_1)
\right]$ is the characteristic function of some set $S_1\subset \R$, 
and similarly for $F_2$, $G_1$ and $G_2$, so that eqs.\rf{7} read
\<\begin{array}{rclrcl}
  r(q_1,q_2) & = & (2\chi_1-1)(2\chi_2-1)\,, &\quad
  s(q_1,p_2) & = & (2\chi_1-1)(2\chi'_2-1)\,, \\
  t(p_1,q_2) & = & (2\chi'_1-1)(2\chi_2-1)\,, &\quad
  u(p_1,p_2) & = & -(2\chi'_1-1)(2\chi'_2-1)\,,
\end{array} \label{12} \>
where $\chi_i$ stands for $\chi_i(q_i)$ and $\chi'_i$ for $\chi'_i(p_i)$,
($i=1,2$). Eqs.\rf{8} then become
\< \P = 0 \mbox{ or } 1 \,, \label{13}
\>
i.e. $\P(1-\P)=0$, where $\P(q_1,q_2,p_1,p_2)$ is given by
\< \P = \chi_1+\chi_2+\chi'_1\chi'_2-\chi_1\chi_2-\chi_1\chi'_2-\chi'_1\chi_2 
   \,. \label{14}
\>
Let us define a corresponding quantum operator $\Ph$ by
\< \Ph = \kh_1+\kh_2+\kh'_1\kh'_2-\kh_1\kh_2-\kh_1\kh'_2-\kh'_1\kh_2 \,,
   \label{15}
\>
where
\<\begin{array}{rclrcl}
  \kh_1 & = & \intd_{S_1}dq_1\,|q_1\rangle\langle q_1|\,\otimes
  \,{\bf 1}_2\,, &\quad
  \kh_2 & = & {\bf 1}_1\,\otimes\,\intd_{S_2}dq_2\,|q_2\rangle
  \langle q_2|\,,\\
  \kh'_1 & = & \intd_{S'_1}dp_1\,|p_1\rangle\langle p_1|\,\otimes
  \,{\bf 1}_2 \,, &\quad
  \kh'_2 & = & {\bf 1}_1\,\otimes\,\intd_{S'_2}dp_2\,|p_2\rangle
  \langle p_2|\,.
\end{array} \label{16} \>
The $\kh$'s are orthogonal projectors,
($\kh\dagger=\kh$,\ $\kh^2=\kh$) acting on ${\cal H}\equiv
L^2(\R,dq_1)\otimes L^2(\R,dq_2)$. The product of two of them involving
different indices commutes, so that $\Ph$ is a (bounded) self-adjoint
operator.

The Bell inequalities \rf{9} to be tested in the quantum context then
become
\< 0\leq\langle\Psi|\Ph|\Psi\rangle\leq1\qquad\forall\ |\Psi\rangle
   \in{\cal H}\mbox{ with }\langle\Psi|\Psi\rangle=1\,, \label{17}
\>
or $0\leq\mbox{Tr}\,\rhoh\,\Ph\leq1$ in case of mixed states. Equivalently,
\< \Ph\geq0 \mbox{\ \ \ and\ \ \ } {\bf 1}-\Ph\geq0
   \mbox{\ \ \ in the operator sense}. \label{18}
\>

	Because $\kh_j$ fails to commute with $\kh'_j$ ($j=1,2$), $\Ph$ is
{\sl not} an orthogonal projector (see below), in contrast to the classical
equality $\P^2=\P$. This leads to the following proposition :

\noindent{\bf The operators $\Ph$ and (${\bf 1}-\Ph$) cannot be both positive.}

\noindent{\sl Proof :} Assume that $\Ph$ and (${\bf 1}-\Ph$) are both
positive. Then
\< \Ph({\bf 1}-\Ph) \geq 0 \,, \label{19}
\>
(remember that the product of two positive {\sl commuting} operators is
positive). 

\noindent Now, a straightforward calculation of $\Ph^2$ yields
\< \Ph^2 = \Ph-\left[\kh_1,\kh'_1\right] \left[\kh_2,\kh'_2\right]\,,
   \label{20}
\>
Take a factorized $|\Psi\rangle$, namely
$|\Psi\rangle=|\Phi_1\rangle\otimes|\Phi_2\rangle$, so that
\< \langle\Psi|\Ph({\bf 1}-\Ph)|\Psi\rangle = 
   -\langle\Phi_1|i\left[\kh_1,\kh'_1\right]|\Phi_1\rangle
   \langle\Phi_2|i\left[\kh_2,\kh'_2\right]|\Phi_2\rangle\,.
   \label{21}
\>
To prove the proposition, it is enough to show that, for a given
choice of the characteristic functions $\chi$ and $\chi'$, the real
number $R[\Phi]\equiv\langle\Phi|i\left[\kh,\kh'\right]|\Phi\rangle$
can assume both signs when $|\Phi\rangle$ is varied. Defining
$|\Phi^+\rangle = \kh\,|\Phi\rangle$ and $|\Phi^-\rangle=(\mbox{\bf
1}-\kh)|\Phi\rangle$, and using the identity
$\left[\kh,\kh'\right]=\kh\kh'({\bf 1}-\kh)-({\bf1}-\kh)\kh'\kh\,$,
gives $R[\Phi]$ the form
\[ R[\Phi] = i\langle\Phi^+|\kh'|\Phi^-\rangle
   -i\langle\Phi^-|\kh'|\Phi^+\rangle\,.
\]
Obviously, for $|\widetilde{\Phi}\rangle=|\Phi^+\rangle-|\Phi^-\rangle$,
one has $R[\widetilde{\Phi}]=-R[\Phi]$. As a consequence,
there is at least one $|\Psi\rangle\neq0$ such that the inequalities
$\langle\Psi|\Ph|\Psi\rangle\geq0$ and
$\langle\Psi|({\bf1}-\Ph)|\Psi\rangle\geq0$ cannot be simultaneously
true, and the four marginal conditions \rf{3} are inconsistent.

\noindent {\sl Remark:}
When the wave function $|\Psi\rangle$ factorizes, i.e.\ 
$\Psi(q_1,q_2) = \Phi_1(q_1)\Phi_2(q_2)$, a corresponding probability
distribution $\rho$ always exists, namely
\[ \rho(q_1,q_2,p_1,p_2) = |\Phi_1(q_1)|^2\,|\Phi_2(q_2)|^2\,
   |\tilde{\Phi}_1(p_1)|^2\,|\tilde{\Phi}_2(p_2)|^2 \,,
\]
where the $\tilde{\Phi}_i$'s are the Fourier transforms
\[ \tilde{\Phi}_i(p_i) = {1\over\sqrt{2\pi}} \int_{-\infty}^{+\infty}dq_i\,
   \mbox{e}^{-ip_iq_i}\,\Phi_i(q_i)\,, \qquad (i=1,2).
\]
Of course, this implies that eqs.\rf{17} are automatically satisfied for
such factorized $|\Psi\rangle$'s (which can also be checked from
eq.\rf{15}).

The above proof is non constructive and gives no quantitative
information about the amount of violation. In order to gain such an
information, one needs to construct explicitely some ``optimal'' wave
function $\psi(q_1,q_2)$, which is not a trivial matter as it involves
a sort of fine tuning. We shall content ourselves to give here the
(surprisingly simple!) result:
\< \Psi_{\pm}(q_1,q_2) = {1\over 2\sqrt{2}}\left[1\pm
   \e{i{\pi\over 4}}\sgn(q_1)\sgn(q_2)\right]h(|q_1|)h(|q_2|)\,,
   \label{22}
\>
where $h(q)$ stands for some regularized form of $1\over\sqrt{q}$, with
$\int_0^\infty dq\,h(q)^2=1$, e.g.
\[ h_L(q) = {\theta(L-q)\over \sqrt{\ln(L+1)}}{1\over \sqrt{q+1}}
   \qquad L\rightarrow\infty.
\]
One can then check that, with the choice
\[ \chi_i(q_i)=\theta(q_i)\,, \quad \chi_i'(p_i)=\theta(p_i)\,, \quad (i=1,2) 
\]
in eq.\rf{12}, the inequalities \rf{9} are violated indeed for
$L\rightarrow\infty$ : $S\rightarrow\pm2\sqrt{2}$.

This opens up the exciting possibility of experimental test of quantum
violation of non contextuality postulate in the phase space context.

\section{General solution of the three marginal problem}
We proved the impossibility of reproducing quantum probabilities of
four CCS as marginals. We now give a sketchy description of the most
general nonnegative phase space density which reproduces any three given
probabilities, say $\sigma_{qq}$, $\sigma_{pq}$ and $\sigma_{pp}$,
satisfying consistency constraints as in eq.\rf{3}. A precise
statement and a full mathematical proof including the required technical
details will be published elsewhere~\cite{AMRS4}.

Let us introduce the one variable marginals
\<\begin{array}{c}
   \displaystyle \sigma_q(q_2) = \int dq_1\,\sigma_{qq}(q_1,q_2)
    = \int dp_1\,\sigma_{pq}(p_1,q_2)\,, \\
   \displaystyle \sigma_p(p_1) = \int dq_2\,\sigma_{pq}(p_1,q_2)
    = \int dp_2\,\sigma_{pp}(p_1,p_2)\,.
\end{array}\label{23}
\>
Let $E=\{\qora,\pora\,|\,\sigma_{qq}, \sigma_{pq}\mbox{ and }
\sigma_{pp}\neq0\}$, and
\< \rho_0(\qp) = \left\{ \begin{array}{l}
     \sigma_{qq}(q_1,q_2) {\displaystyle 1\over\displaystyle\sigma_q(q_2)}
     \sigma_{pq}(p_1,q_2) {\displaystyle1\over\displaystyle\sigma_p(p_1)}
     \sigma_{pp}(p_1,p_2) \mbox{\ \ if\ }(\qp)\in E\,, \\
     0 \qquad \mbox{ otherwise.} \end{array}\right. \label{24}
\>
Clearly $\rho_0$ is a particular non negative solution of the given
three marginal constraints. We now state the theorem :

{\bf The general nonnegative $\rho(\qp)$ with prescribed marginals
$\sigma_{qq}$, $\sigma_{pq}$ and $\sigma_{pp}$ is given by
\< \rho(\qp) = \rho_0(\qp)+\lambda\,\Delta(\qp)\,, \label{25}
\>
where
\< \lambda\in\left[-{1/m_+},{1/m_-}\right]\,, \label{26}
\>
and
\renewcommand{\arraystretch}{1.9}
\<\begin{array}{l}
  \Delta(\qp) = F(\qp)-\rho_0(\qp)
  \left[\displaystyle
  {1\over\sigma_{qq}(q_1,q_2)}\intd dp'_1dp'_2
  \,F(q_1,q_2,p'_1,p'_2) \right. \\
   \qquad +\displaystyle
  {1\over\sigma_{pq}(p_1,q_2)}\intd dq'_1dp'_2
  \,F(q'_1,q_2,p_1,p'_2) +
  {1\over\sigma_{pp}(p_1,p_2)}\intd dq'_1dq'_2
  \,F(q'_1,q'_2,p_1,p_2) \\
   \qquad -\displaystyle\left.
  {1\over\sigma_q(q_2)}\intd dq'_1dp'_1dp'_2
  \,F(q'_1,q_2,p'_1,p'_2) -
  {1\over\sigma_p(p_1)}\intd dq'_1dq'_2dp'_2
  \,F(q'_1,q'_2,p_1,p'_2) \right]\,,
  \end{array}\label{27}
\>
\renewcommand{\arraystretch}{1.4}
$F$ being an {\Bxsl arbitrary} function with support contained in
$E$. The ($F$-dep\-end\-ent) constants $m_\pm$ in \rf{26} are defined as
\< m_+ = \ \build{\mbox{\rm sup }}{(\overrightarrow{\scriptstyle q},
   \overrightarrow{\scriptstyle p})\in E} \ {\Delta(\qp)\over\rho_0(\qp)}\,,
   \qquad
   m_- = \ -\build{\mbox{\rm inf }}{(\overrightarrow{\scriptstyle q},
   \overrightarrow{\scriptstyle p})\in E} \ {\Delta(\qp)\over\rho_0(\qp)}\,,
   \label{28}
\>
and are both positive if $\Delta$ does not identically vanish
($m_+=\infty$ or/and $m_-=\infty$ are not excluded).}

The proof goes in two steps. First, it is readily
shown that any non negative solution $\rho_1$ of the three marginal
conditions admits the representation \rf{25}. Indeed, choosing
$F=\rho_1$ in eq.\rf{27} gives $\Delta=\rho_1-\rho_0$ and $m_-\leq1$,
allowing to choose $\lambda=1$. Eq.\rf{25} then reads $\rho=\rho_1$.

Second, one shows that any function $\rho$ defined by \rf{25} to
\rf{28} is a non negative solution of the three marginal
conditions. To do it, it is convenient to rearrange the writing of
$\Delta$ as follows~:
\begin{eqnarray}
  \Delta &=& \left[F-{\rho_0\over\sigma_{qq}} \int dp'_1dp'_2\,F\right]
  -\left[{\rho_0\over\sigma_{pq}} \int dq'_1dp'_2\,F
  - {\rho_0\over\sigma_q} \int dq'_1dp'_1dp'_2\,F\right] \nonumber \\
  && -\left[{\rho_0\over\sigma_{pp}} \int dq'_1dq'_2\,F
  - {\rho_0\over\sigma_p} \int dq'_1dq'_2dp'_2\,F\right]\,. \label{29}
\end{eqnarray}
Integrating the right-hand side over $p_1$ and $p_2$, one finds that the two terms
coming from each square bracket cancel each other. Similar results
obtain on integrating over $(q_1,p_2)$ or $(q_1,q_2)$. Hence
\[ \left\{\int dp_1dp_2,\int dq_1dp_2,\int dq_1dq_2\right\}\,\Delta(\qp) = 0\,,
\]
and the three marginal conditions are satisfied by \rf{25}.

Since the integral of $\Delta$ over phase space vanishes, $m_\pm$ in
eqs.\rf{28} are both strictly positive if $\Delta$ does not vanish
identically. The positivity of $\rho$ then follows from eqs.\rf{25},
\rf{26} and \rf{28}.

Finally, combining the proposition of section IV with the above
theorem, we can state the

\noindent{\bf Three marginal theorem :} {\sl Any three out of a given
set of four probability densities obeying the consistency conditions
\rf{5} can be reproduced as marginals of a positive density
$\rho(\qp)$. There exist sets of four consistent probability densities
which cannot be reproduced as marginals of a positive $\rho$.}

\section{Conclusions}
We established phase space Bell inequalities from the postulate of
existence of a positive phase space probability density. We
demonstrated that quantum mechanics violates these inequalities by a
factor $\sqrt{2}$, as in the violation of the standard ones, opening
the road to experimental tests of quantum contextuality in the
position-momentum sector. We also established the three marginal
theorem which shows that in 2-dimensional configuration space, three
(but not four) noncommuting CCS can be simultaneously realized in quantum
mechanics. The simultaneous realization of three CCS (rather than the
usual 1 CCS) and the construction of the most general such phase space
density sets the stage for construction of maximally realistic quantum theory.

\section{Acknowledgements}

     We thank Andr\'e Martin for collaboration in the initial stages of
this work.  One of us (SMR) thanks A. Fine and A. Garg for some
remarks on the three marginal problem many years ago.

\end{document}